\documentclass{kyriakos}

\newcommand{\Qx}{ \mathbb{Q} }
\newcommand{\Ex}{ \mathbb{E} }

\newcommand{\rec}{\mbox{R{\tiny EC}}}

\newcommand{\lgd}{\mbox{L{\tiny GD}}}

\newcommand{\npv}{\mbox{NPV}}

\newcommand{\cds}{\mbox{CDS}}

\newcommand{\PCS}{\mbox{PCS}}
\newcommand{\RCS}{\mbox{RCS}}

\newtheorem{theorem}{Theorem}[section]

\newtheorem{proposition}[theorem]{Proposition}

\numberwithin{equation}{section}

\title{Counterparty risk valuation for Energy-Commodities swaps}
\shorttitle{Counterparty risk for Energy-Commodities swaps}
\subtitle{{\bf Impact of volatilities and correlation}}
\author{Damiano Brigo  \\
Fitch Solutions\\ and Dept. of Mathematics \\ Imperial College\\
\and Kyriakos Chourdakis \\
Fitch Solutions\\ and CCFEA \\ Univ. of Essex \\ \and Imane Bakkar \\ Fitch Solutions \\ 101 Finsbury Pavement \\ London
}
\shortauthor{Brigo, D., Bakkar, I., and Chourdakis, K.}

\usepackage[absolute]{textpos}
\usepackage{epsfig}

\begin{document}

%\%begin{textblock}{8}(1.6,1)
%%PNG_LOGO
%\includegraphics[width=10cm]{FSLogo_.PNG}
%\end{textblock}

\maketitle \thispagestyle{empty}

\begin{abstract}
It is commonly accepted that Commodities futures and forward prices, in principle, agree under some simplifying assumptions. One of the most relevant assumptions is the absence of counterparty risk. Indeed, due to margining, futures have practically no counterparty risk. Forwards, instead, may bear the full risk of default for the counterparty when traded with brokers or outside clearing houses, or when embedded in other contracts such as swaps.
In this paper we focus on energy commodities and on Oil in particular. We use a hybrid commodities-credit model to asses impact of counterparty risk in pricing formulas, both in the gross effect of default probabilities and on the subtler effects of credit spread volatility, commodities volatility and credit-commodities correlation.
We illustrate our general approach with a case study based on an oil swap, showing that an accurate valuation of counterparty risk depends on volatilities and correlation and cannot be accounted for precisely through a pre-defined multiplier.
\end{abstract}

\medskip

\noindent {\bf AMS Classification Codes}: 60H10, 60J60, 60J75, 62H20, 91B70


\bigskip


\noindent Oil models, Convenience Yield models, Stochastic Intensity models.

\bigskip

\noindent Fitch Solutions, 101 Finsbury Pavement, EC2A 1RS London

\noindent E-mail:

 {\tt damiano.brigo@fitchsolutions.com}

\medskip

\noindent {June 24, 2008. Available also at www.damianobrigo.it and at SSRN.com }

\newpage

\tableofcontents

\newpage

\section{Introduction}

In this paper we consider counterparty risk for commodities
payoffs in presence of correlation between the default event and
the underlying commodity, while taking into account volatilities both for credit and commodities.
We focus on Oil but much of our reasoning can be adapted to other commodities with similar characteristics (storability, liquidity, and similar seasonality).

Past work on {\em pricing} counterparty risk for different asset classes is in Sorensen and Bollier (1994),
Brigo and Masetti (2006) and Brigo and Pallavicini (2007, 2008) for interest rate swaps and exotics
underlyings. Leung and Kwok (2005) and Brigo and Chourdakis (2008) worked on counterparty risk for credit (CDS) underlyings.

Here we analyze in detail counterparty-risky (or
default-risky) Oil forward and swaps contracts.

In general the reason to introduce counterparty risk when
evaluating a contract is linked to the fact that many financial
contracts are traded over the counter, so that the credit quality
of the counterparty can be relevant. This is particularly
appropriated when thinking of the different defaults experienced
by some important companies during the last years, especially in the energy sector.

Earlier works in counterparty risk for commodities include for example Cannabaro, Picoult and Wilde (2005), who analyze this notion more from a capital adequacy/ risk management point of view. In particular, their approach is not dynamical and does not consider explicitly credit spread volatility and especially correlation between the underlying commodity and credit spread. In our approach wrong way risk is modeled through said correlation. Mostly, however, the difference is in the purpose. We are valuing counterparty risk more from a pricing than a risk management perspective, resorting to a fully arbitrage free and fine-tuned risk neutral approach. This is why all our processes are calibrated to liquid market information both on forward curves and volatilities. Correlations are harder to estimate but we analyze their impact by letting them range across a set of possible values. A limitation of our approach is that we refer to a single counterparty.
%Our focus is in valuing accurately the counterparty risk price for a single deal rather than getting the impact %roughly right for a very large portfolio.

In general we are looking at the problem from the viewpoint of a safe
(default-free) institution entering a financial contract with
another counterparty having a positive probability of defaulting
before the final maturity. We formalize the general and reasonable
fact that the value of a generic claim subject to counterparty
risk is always smaller than the value of a similar claim having a
null default probability, expressing the discrepancy in precise
quantitative terms.

We consider Credit Default Swaps for the counterparty as liquid sources of market default probabilities.
Different models can be used to calibrate CDS data and obtain
default probabilities: here we resort to
Brigo and Alfonsi (2005) stochastic intensity model, whose jump extension with analytical formulas for CDS options is illustrated in Brigo and El-Bachir (2008).

As a model for oil we adopt a two factor model shaping both the short term deviation in prices and the equilibrium price level, as in Smith and Schwartz (2000). This model can be shown to be equivalent to a more classical convenience yield model like in Gibson and Schwartz (1990), and a stochastic volatility extension of a similar approach is considered in Geman (2000). What is modeled is the oil spot price, under the implicit assumption that such a spot price process exists. This is not true for electricity, for example, and even for
markets like crude oil where spot prices are quoted daily, the exact meaning of the spot
is difficult to single out. Nonetheless, we assume, along with most of the industry and with Carmona and Ludkowski (2004), that there is a traded spot asset.

In the paper we find that counterparty risk has a relevant impact on the products prices and that, in turn, correlation between oil and credit spreads of the counterparty has a relevant impact on the adjustment
due to counterparty risk. Similarly, oil and credit spread volatilities have reasonable impacts on the adjustment. The impact patterns do not involve the peculiar behaviour one observes in the case of credit underlyings, observed in Brigo and Chourdakis (2008). Nonetheless, the impact is quantitatively relevant, and we illustrate this with a case study based on an oil swap.

The paper is organized as follows: Section~\ref{sec:generalcounterformula} lays down the general framework for the valuation of counterparty risk. In section~\ref{sec:assumptions_credit} we present the CIR++ specification which serves as the credit model, and in section~\ref{sec:assumptions_comm} we outline the two-factor Smith and Schwartz commodity model. Sections~\ref{sec:for_fut} and~\ref{sec:swap} illustrate the counterparty adjustments for forwards and swaps respectively. An example, based on a swap contract with between a bank and an airline company is presented in section~\ref{sec:empirical}.

\section{General valuation of counterparty risk}\label{sec:generalcounterformula}

We denote by $\tau$ the default time of the counterparty and we
assume the investor who is considering a transaction with the
counterparty to be default-free. We place ourselves in a
probability space $(\Omega,\mathcal{G},\mathcal{G}_t,\mathbb{Q})$.
The filtration $(\mathcal{G}_t)_t$ models the flow of information
of the whole market, including credit and defaults. $\Qx$ is the risk neutral
measure. This space is endowed also with a right-continuous and
complete sub-filtration $\mathcal{F}_t$ representing all the
observable market quantities but the default event (hence
$\mathcal{F}_t\subseteq\mathcal{G}_t:=\mathcal{F}_t\vee\mathcal{H}_t$
where $\mathcal{H}_t=\sigma(\{\tau\leq u\}:u\leq t)$ is the
right-continuous filtration generated by the default event). We
set $\mathbb{E}_t(\cdot):=\mathbb{E}(\cdot|\mathcal{G}_t)$, the
risk neutral expectation leading to prices.

Let us call $T$ the final maturity of the payoff we need to
evaluate. If $\tau>T$ there is no default of the counterparty
during the life of the product and the counterparty has no
problems in repaying the investors. On the contrary, if $\tau\leq
T$ the counterparty cannot fulfill its obligations and the
following happens. At $\tau$ the Net Present Value (NPV) of the
residual payoff until maturity is computed: If this NPV is
negative (respectively positive) for the investor (defaulted
counterparty), it is completely paid (received) by the investor
(counterparty) itself. If the NPV is positive (negative) for the
investor (counterparty), only a recovery fraction $\rec$ of the
NPV is exchanged.

Let us call $\Pi^D(t,T)$ (sometimes abbreviated into $\Pi^D(t)$)
the discounted payoff of a generic claim at $t$ under counterparty risk. This is the sum of all cash flows from $t$ to $T$, each discounted back at $t$, and under counterparty risk. This is a stochastic payoff, whose price would be given by risk neutral expectation. We denote by $\Pi(t,T)$ the analogous quantity when counterparty risk is absent, or when the counterparty is default free. All payoffs
are seen from the point of view of the ``investor" (i.e. the company
facing counterparty risk). Then we have
$\npv(\tau)=\mathbb{E}_{\tau}\{\Pi(\tau,T)\}$ and
\begin{eqnarray}\label{generalpayoff}
\nonumber \Pi^D(t) & = & \mathbf{1}_{\{\tau>T\}} \Pi(t,T) + \\
& & \mathbf{1}_{\{t < \tau\leq T\}}
\left[\Pi(t,\tau)+D(t,\tau)\left(\rec\left(\npv(\tau)\right)^+-\left(-\npv(\tau)\right)^+\right)\right]
\end{eqnarray}
being $D(u,v)$ the stochastic discount factor at time $u$ for
maturity $v$. This last expression is the general price of the
payoff under counterparty risk. Indeed, if there is no early
counterparty default this expression reduces to risk neutral valuation of the
payoff (first term in the right hand side); in case of early
default, the payments due before default occurs are received
(second term), and then if the residual net present value is
positive only a recovery of it is received (third term), whereas
if it is negative it is paid in full (fourth term).

Calling $\Pi(t)$ the discounted payoff for an equivalent claim
with a default-free counterparty, i.e. $\Pi(t) = \Pi(t,T)$, it
is possible to prove the following
\begin{proposition} {\bf (General counterparty-risk credit-valuation adjustment (CR-CVA)
formula)}. At valuation time $t$, and on $\{ \tau > t\}$, the
price of our payoff under counterparty risk is
\begin{eqnarray}\label{generalprice}
\mathbb{E}_t\{\Pi^D(t)\} =
\mathbb{E}_t\{\Pi(t)\}-&\underbrace{\lgd\,\,\mathbb{E}_t\{\mathbf{1}_{\{t
< \tau\leq T\}}D(t,\tau)\left(\npv(\tau)\right)^+}\}&\\ \nonumber
&\mbox{Positive CR-CVA}&
\end{eqnarray}
where $\lgd=1-\rec$ is the \emph{Loss Given Default} and the
recovery fraction $\rec$ is assumed to be deterministic. It is
clear that the value of a defaultable claim is the value of the
corresponding default-free claim minus an option part, in the
specific a call option (with zero strike) on the residual NPV
giving nonzero contribution only in scenarios where $\tau\leq T$.
Counterparty risk adds an optionality level to the original
payoff.
\end{proposition}
For a proof see for example Brigo and Masetti (2006).

Notice finally that the previous formula can be approximated as
follows. Take $t=0$ for simplicity and write, on a discretization
time grid $T_0,T_1,\ldots, T_b=T$,

\begin{eqnarray}\label{eq:adjustment_discretized}
\nonumber \Ex[\Pi^D(0,T_b)] = \Ex[\Pi(0,T_b)] -  &\lgd
\sum_{j=1}^b \Ex[ \mathbf{1}_{\{T_{j-1} < \tau \le T_j\}}D(0,\tau)
(\Ex_{\tau} \Pi(\tau,T_b))^+]&\\
 \approx \Ex[\Pi(0,T_b)] -  &\underbrace{\lgd \sum_{j=1}^b
\Ex[\mathbf{1}_{\{T_{j-1} < \tau \le T_j\}}D(0,T_j) (\Ex_{T_j}
\Pi(T_j,T_b))^+]}&\\ \nonumber &\mbox{approximated (positive)
adjustment}
 \end{eqnarray}
where the approximation consists in postponing the default time to
the first $T_i$ following $\tau$. From this last expression, under
independence between $\Pi$ and $\tau$, one can factor the outer
expectation inside the summation in products of default
probabilities times option prices. This way we would not need a
default model for the counterparty but only survival probabilities and an option model
for the underling market of $\Pi$. This is what led to earlier results on swaps with counterparty risk in
interest rate payoffs in Brigo and Masetti (2006). In this paper we do not assume zero
correlation, so that in general we need to compute the
counterparty risk without factoring the expectations.

\section{Default modeling assumptions}\label{sec:assumptions_credit}

In this section we consider a reduced form model that is stochastic in the default intensity for the
counterparty. We will later correlate the credit spread of this model with the underlying commodity model.

More in detail, we assume that the counterparty default intensity
is $\lambda$, and we denote the cumulated intensity by $\Lambda(t) = \int_0^t \lambda(s)ds$. We assume intensities to be strictly positive, so that $t \mapsto \Lambda$ are invertible functions.

We assume deterministic default-free instantaneous interest rate $r$ (and hence deterministic discount factors $D(s,t),...$), although our analysis would work well even with stochastic rates independent of oil and credit spreads.

We set ourselves in a  Cox process setting, where
\[ \tau = \Lambda^{-1}(\xi), \]
with $\xi$ standard (unit-mean) exponential random variable.

\subsection{CIR++ stochastic intensity  models}

For the stochastic intensity model we set
\begin{equation} \label{extended_i}
\lambda(t) = y(t) + \psi(t;\beta)\ , \ \ t\geq 0,
\end{equation}
where $\psi$ is a deterministic function, depending on the
parameter vector $\beta$ (which includes $y_0$), that is
integrable on closed intervals. The initial condition $y_0$ is one
more parameter at our disposal: We are free to select its value as
long as
\[ \psi(0;\beta) = \lambda_0 - y_0  \ . \]
We take $y$ to be a Cox Ingersoll Ross process (see for example
Brigo and Mercurio (2001) or (2006)):
\[dy(t)= \kappa(\mu-y(t))dt+\nu \sqrt{y(t)}dZ_y(t),\]
where the parameter vector is $\beta = (\kappa, \mu, \nu,y_0)$,
with $\kappa$, $\mu$, $\nu, y_0$ positive deterministic constants.
As usual,  $Z_y$ is a standard Brownian motion processes under the risk
neutral measure, representing the stochastic shock in our
dynamics. We assume the origin to be inaccessible, i.e.
\[ 2 \kappa \mu > \nu^2 .  \]
%For restrictions on the $\beta$'s that keep $\lambda$ positive, as
%is required in intensity models, refer to Brigo and Mercurio (2001,2006).

We will often use the integrated quantities \[ \Lambda(t)=\int_0^t
\lambda_s ds,\ \  Y(t)=\int_0^t y_s ds, \ \ \mbox{ and} \ \
\Psi(t,\beta)=\int_0^t \psi(s,\beta)ds.\]

%\subsection{Calibrating the joint stochastic model to CDS}\label{calsepar}
\subsection{CIR++ model: CDS calibration}

Since we are assuming deterministic rates,
the default time $\tau$ and interest rate
quantities $r, D(s,t),...$ are trivially independent. It follows that the (receiver) CDS valuation at time $0$ becomes model independent and is given by the formula
%[insert detailed description of CDS contract]
%
\begin{eqnarray}\label{ch:credit:modindcdstot}
\hspace{-1cm} \cds_{a,b}(0, S_{a,b}, \lgd;\Qx(\tau > \cdot))  =  S_{a,b} \left[ -\int_{T_a}^{T_b} P(0,t) (t-T_{\gamma(t)-1})
d_t \boxed{\Qx(\tau \ge t)}\right. \\ + \left.
   \sum_{i=a+1}^b P(0,T_i)  \alpha_i  \boxed{ \Qx(\tau \ge
 T_i)} \right] +\\ + \lgd \left[
\int_{T_a}^{T_b} P(0,t)\  d_t \boxed{\Qx(\tau \ge t)}\right]
\end{eqnarray}
%(see for example the Credit chapters in Brigo and Mercurio
%(2006)).
This means that if we strip survival probabilities from
CDS in a model independent way at time 0, to calibrate the market CDS quotes
we just need to make sure that the survival probabilities we strip
from CDS are correctly reproduced by the CIR++ model. Since the
survival probabilities in the CIR++ model are given by
\begin{equation}\label{ch:credit:cdscalrho0} \mathbb{Q}(\tau>t)_{\mbox \tiny
model}=\mathbb{E}(e^{-\lambda(t)}) =  \ \mathbb{E}
\exp\left(-\Psi(t,\beta) - Y(t)\right)
\end{equation}
we just need to make sure
\begin{equation*}
\ \mathbb{E} \exp\left(-\Psi(t,\beta) - Y(t)\right) =
\mathbb{Q}(\tau>t)_{\mbox \tiny market}
\end{equation*}
%If we agree to express survival probabilities through implied
%hazard rates and functions $\Gamimpl$, defined as $\Qx(\tau >
%t)_{\mbox \tiny market} = e^{-\Gamimpl(t)}$, then
%Equation~(\ref{ch:credit:cdscalrho0}) reads
%\begin{equation}\label{ch:credit:cdscalrho0bis}
% \Bbb{E} \exp\left(-\Psi(t,\beta) -
%Y^\beta(t)\right) = e^{-\Gamimpl(t)}
%\end{equation}
from which
\begin{equation}\label{fittingPsicir}    \Psi(t,\beta)=
\ln\left(\frac{\mathbb{E}(e^{-Y(t)})}{\mathbb{Q}(\tau>t)_{\mbox \tiny
market}}\right) = \ln\left(\frac{P^{\mbox{\tiny CIR}}(0,t, y(0);
\beta)}{\mathbb{Q}(\tau>t)_{\mbox \tiny market}}\right)
\end{equation}
where we choose the parameters $\beta$ in order to have a positive
function $\psi$ (i.e. an increasing $\Psi$) and $P^{\mbox{\tiny
CIR}}$ is the closed form expression for bond prices in the time
homogeneous CIR model with initial condition $y_0$ and parameters
$\beta$ (see for example Brigo and Mercurio (2001, 2006)). Thus,
if $\psi$ is selected according to this last formula, as we will
assume from now on, the model is easily and automatically
calibrated to the market survival probabilities for the counterparty (possibly stripped
from CDS data).

Once we have done this and calibrated CDS data through
$\psi(\cdot,\beta)$, we are left with the parameters $\beta$,
which can be used to calibrate further products. However, this
will be interesting when single name option data on the credit
derivatives market will become more liquid. Currently the bid-ask
spreads for single name CDS options are large and suggest to
either consider these quotes with caution, or to try and deduce volatility parameters from more liquid index options. At the moment we content
ourselves of calibrating only CDS's. To help
specifying $\beta$ without further data we set some values of the
parameters implying possibly reasonable values for the implied
volatility of hypothetical CDS options on the counterparty.

%In our tests we take stylized flat CDS curves for the
%counterparty, assuming they imply initial survival probabilities
%at time $0$ consistent with the following hazard function
%formulation,
%\[   \Qx(\tau > t)_{\mbox{\tiny market}} = \exp(-\gamma t), \]
%%
%for a constant deterministic value of $\gamma$. This is to be
%interpreted as a quoting mechanism for survival probabilities and
%not as a model. Assuming our counterparty CDS's at time $0$ for
%different maturities  to imply a given value of $\gamma$, we will
%value counterparty risk under different values of $\gamma$. This
%assumption on CDS spreads is stylized but our aim is checking
%impacts rather than having an extremely precise valuation.
%
%In our numerical examples we take as values of the intensity
%volatility parameters $y_0, \kappa, \mu, \nu$ the following
%values:
%\[ y_0=0.0165,\; \kappa=0.4,\; \mu=0.026,\; \nu=0.14 \]
%%
%Paired with stylized CDS data consistent with survivals $\Qx(\tau
%> t)_{\mbox{\tiny market}} = \exp(-\gamma t)$ for several possible
%values of $\gamma$, these parameters imply the CDS
%volatilities\footnote{See Brigo (2005, 2006) for a precise notion
%of CDS implied volatility.} listed in Table \ref{implvol}.
%
%% , where we have assumed deterministic
%%interest rates in determining CDS implied volatilities (the impact of the interest rate
%%volatility on the CDS payoff is very limited).

\section{Commodity model}\label{sec:assumptions_comm}
We consider crude oil as a first important case.

Suppose we have a airline company that buys a forward contract on oil from a bank with a very high credit quality, so that we assume the bank to be default-free. The bank wants to charge counterparty risk to the airline in defining the forward price, as there is no collateral posted and no margining is occurring.

As a model for oil we adopt a two factor model shaping both the short term deviation in prices and the equilibrium price level, as in Smith and Schwartz (2000). This model can be shown to be equivalent to a more classical convenience yield model like in Gibson and Schwartz (1990), and a stochastic volatility extension of a similar approach is considered in Geman (2000). What is modeled is the oil spot price, under the implicit assumption that such a spot price process exists. This is not true for electricity, for example, and even for
markets like crude oil where spot prices are quoted daily, the exact meaning of the spot
is difficult to single out. Nonetheless, we assume, along with most of the industry, that there is a traded spot asset.

If we denote by $S_t$ the oil spot price at time $t$, the log-price process is written as
\[ \ln(S_t) = x(t) + L(t) + \varphi(t) , \]
where, under the risk neutral measure,
\begin{eqnarray}
d x(t) &=& - k_x x(t) dt + \sigma_x d Z_x,\\
d L(t) &=& \mu_L dt + \sigma_L dZ_L, \ \ \  dZ_x \ dZ_L = \rho_{x,L} dt,
\end{eqnarray}
and $\varphi$ is a deterministic shift we will use to calibrate quoted futures prices. The process $x$ represents the short term deviation, whereas $L$ represents the backbone of the equilibrium price level in the long run.

For applications it can be important to derive the transition density of the spot commodity in this model. For the two factors we have a joint Gaussian transition,
\begin{eqnarray*}
\left[\begin{array}{c} x(t) \\ L(t) \end{array} \right] \bigg{|}_{x(s),L(s)} \sim {\cal N}
\left( \left[\begin{array}{c} x(s) \exp(- k_x (t-s)) \\ L(s) + \mu_L (t-s) \end{array} \right] , \left[ \begin{array}{cc} \frac{\sigma_x^2}{2 k_x }(1-\exp(- 2 k_x (t-s))) & \mbox{Cov}_{x,L}(s,t)  \\ &  \sigma_L^2 (t-s) \end{array}  \right] \right) \\
 \mbox{Cov}_{x,L}(s,t)  = \rho_{x,L} \frac{\sigma_x \sigma_L}{ k_x }(1-\exp(- k_x (t-s))) .
\end{eqnarray*}

This can be used for exact simulation between times $s$ and $u$.  As we know that the sum of two jointly Gaussian random variables is Gaussian, we have
\begin{eqnarray*}
\ln(S(t)) = x(t) + L(t)+\varphi(t)|_{x(s),L(s)} \sim {\cal N}( m(t,s) , V(s,t) )\\
m(t,s) = x(s) \exp(- k_x (t-s))+ L(s) + \mu_L (t-s) + \varphi(t),\\
V(t,s) = \frac{\sigma_x^2}{2 k_x }(1-\exp(- 2 k_x (t-s))) + \sigma_L^2 (t-s) + 2 \mbox{Cov}_{x,L}(s,t)
\end{eqnarray*}
from which, in particular, we see that
\begin{eqnarray*}
\Ex [ S(t) | x(s),L(s)] = \exp(x(s) \exp(- k_x (t-s))+ L(s) + \mu_L (t-s) + \varphi(t)+ V(s,t)/2)
\end{eqnarray*}

Hence we can compute the forward price $\Ex [ S(T) | x(t),L(t)]$ at time $t$ of the commodity at maturity $T$ when counterparty risk is negligible and under deterministic interest rates, as

\begin{eqnarray}\label{oil:fwd:smith}
F(t,T) =  \exp(x(t) \exp(- k_x (T-t))+ L(t) + \mu_L (T-t) + \varphi(T) + V(T,t)/2)
\end{eqnarray}

In particular, given the forward curve $T \mapsto F^M(0,T)$ from the market, the expression for the shift $\varphi^M(T)$ that makes the model consistent with said curve is

\[ \varphi^M(T) = \ln (F^M(0,T)) - x_0 \exp(-k_x T) - L_0 - \mu_L T - V(T,t)/2 . \]

The short term/equilibrium price model $(x,L)$, when $\varphi = 0$, is equivalent to the more classical Gibson and Schwartz (1990) model, formulated as

\begin{eqnarray}
d \ln(S_t) &=& (r(t)- q(t) -  \sigma_S^2/2 )dt + \sigma_S d Z_S,\\ \nonumber
d q(t) &=& k_q (\alpha - q(t)) dt + \sigma_q dZ_q, \ \ \  dZ_S \ dZ_q = \rho_{q,S} dt,
\end{eqnarray}
the relationships being
\begin{eqnarray*}
 x(t) &=& \frac{1}{k_q} (q(t)-\alpha) \\
 L(t) &=& \ln(S_t) - \frac{1}{k_q} (q(t)-\alpha)\\
k_x &=& k_q \\
\mu_L &=& r - \alpha - \sigma_S^2/2 \\
\sigma_x &=& \sigma_q / k_q\\
\sigma_L^2 &=& \sigma_S^2 + \sigma_q^2 / k_q^2 - 2 \rho_{q,S} \sigma_S \sigma_q / k_q \\
d Z_x &=& d Z_q \\
d Z_L &=& \left(\sigma_S dZ_S - \frac{\sigma_q }{k_q} d Z_q\right) / \sqrt{ \sigma_S^2 + \sigma_q^2 / k_q^2 - 2 \rho_{q,S} \sigma_S \sigma_q / k_q} \\
\rho_{x,L} &=& \left(\sigma_S \rho_{q,s} - \frac{\sigma_q }{k_q} \right) / \sqrt{ \sigma_S^2 + \sigma_q^2 / k_q^2 - 2 \rho_{q,S} \sigma_S \sigma_q / k_q}
\end{eqnarray*}

\section{Forward vs Future prices and counterparty risk}\label{sec:for_fut}

Consider now a forward contract. The propotypical forward contract agrees on the following.

Let $t$ be the valuation time. At the future time $T$ a party agrees to buy from a second party a commodity at the price $K$ fixed today. This is expressed by saying that the first party has entered a payer forward rate agreement. The second party has agreed to enter a receiver forward rate agreement. The value of this contract to the first and second party respectively, at maturity, will be
\[ S_T - K , \ \  K - S_T\]
i.e. the actual price of the commodity at maturity minus the pre-agreed price in the payer case, and the opposite of this in the receiver case. Let us focus on the payer case.
When this is discounted back at $t$ with deterministic interest rates, and risk neutral expectation is taken, this leads to the price being given by
\begin{eqnarray}\label{fw:contract:price} \Ex_t[D(t,T)( S_T - K) ] = D(t,T) (\Ex_t[S_T] - K) = D(t,T) (F(t,T) - K) .
\end{eqnarray}
Note that the forward price is exactly the value of the pre-agreed rate $K$ that sets the contract price to zero, i.e. $K=F(t,T)$. Let us maintain a general $K$ in the forward contract under examination.

In the oil model above, the forward contract price is given by plugging Formula~(\ref{oil:fwd:smith}) into
(\ref{fw:contract:price}). Let us denote by Fwdp$(t,T;K)$ such price (``p" is for payer),

\[ \mbox{Fwdp}(t,T;K)= D(t,T) \left(  \exp(x(t) \exp(- k_x (T-t))+ L(t) + \mu_L (T-t) + \varphi(T) + V(T,t)/2) - K \right) \]
whereas the opposite of this quantity is denoted by Fwdr$(t,T;K)$.

We may apply our counterparty risk framework to the forward contract, where now $\Pi(t,T)=D(t,T)( S_T - K)$, and $NPV(t) = $Fwdp$(t,T;K)$. We obtain as price of the payer forward under counterparty risk from Equation~(\ref{generalprice}). We obtain
\begin{eqnarray}\label{generalprice:fwdcom}
\mbox{Fwdp}^D(t,T;K) = \mbox{Fwdp}(t,T;K)
-&\underbrace{\lgd\,\,\mathbb{E}_t\{\mathbf{1}_{\{t
< \tau\leq T\}}D(t,\tau)\left(\mbox{Fwdp}(\tau,T;K) \right)^+}\}&\\ \nonumber
&\mbox{Positive counterparty-risk adjustment}&
\end{eqnarray}
Under the bucketing approximation given by Equation~(\ref{eq:adjustment_discretized}), we obtain

\[ \mbox{Fwdp}^D(t,T;K)= \mbox{Fwdp}(t,T;K) - \lgd \sum_{j=1}^b D(t,T_j)
\Ex_t[\mathbf{1}_{\{T_{j-1} < \tau \le T_j\}} ( \mbox{Fwdp}(T_j,T;K)  )^+]. \]
If one assumes independence between the underlying commodity and the conterparty default, one may factor the above expectation obtaining
\[ \mbox{Fwdp}^D(t,T;K)= \mbox{Fwdp}(t,T;K) - \lgd \sum_{j=1}^b
\Qx(\{T_{j-1} < \tau \le T_j\})\Ex_t[D(t,T_j)( \mbox{Fwdp}(T_j,T;K)  )^+]. \]
The last term is the price of a option on a forward price, that is known in closed form in the Schwartz and Smith model, although we have to incorporate the shift in our formulation. We have
\begin{eqnarray*}
&&\Ex_t[D(t,T_j)( \mbox{Fwdp}(T_j,T;K)  )^+] = \\
&=& D(t,T)\exp(M(t,T,T_j;x_t,L_t)+\bar{V}(t,T,T_j)/2)
\Phi\left(\frac{M(t,T,T_j;x_t,L_t)+\bar{V}(t,T,T_j)-\ln K}{\sqrt{\bar{V}(t,T,T_j)}}   \right)\\
&& - D(t,T) K \Phi\left(\frac{M(t,T,T_j;x_t,L_t)-\ln K}{\sqrt{\bar{V}(t,T,T_j)}}  \right)\\
&& M(t,T,T_j;x_t,L_t) = x_t \exp(-k_x (T-t)) + L_t + \varphi(T) + V(T,T_j)/2\\
&& \bar{V}(t,T,T_j) = \exp(- 2 k_x (T-T_j)) \frac{\sigma_x^2}{2 k_x }(1-\exp(- 2 k_x (T_j-t)))\\
&& + \sigma_L^2 (T_j-t) + \exp(-  k_x (T-T_j)) 2 \rho_{x,L} \frac{\sigma_x \sigma_L}{ k_x }(1-\exp(- k_x (T_j-t)))
\end{eqnarray*}

where $\Phi$ is the cumulative distribution function of the standard Gaussian.

So we have the adjustment as a stream of options on forwards weighted by default probabilities.

If we do not assume independence then we need to substitute for the intensity model. Through iterated conditioning we obtain easily
\[ \mbox{Fwdp}^D(t,T;K)= \mbox{Fwdp}(t,T;K)\]\[ - \lgd \sum_{j=1}^b D(t,T_j)
\Ex_t[  (\exp(-\Lambda(T_{j-1})) - \exp(-\Lambda(T_{j}) ) )( \mbox{Fwdp}(T_j,T;K;x(T_j),L(T_j))  )^+]. \]
If in particular we select $K = F(t,T)$ then $\mbox{Fwdp}(t,T;K)$ will be zero.

This price can be computed by joint simulation of $\lambda$, $x$ and $L$. We may correlate the credit spread to the commodity by correlating the shock $Z_y$ in the default intensity to the shocks $Z_x, Z_L$ in the commodity.
If we assume

\[ d Z_x d Z_y = \rho_{x,y} dt, \ \  d Z_L d Z_y = \rho_{y,L} dt \]

then the instantaneous correlation of interest is

\[ \mbox{corr}(d \lambda_t , \ d S_t) = \frac{\sigma_x \rho_{x,y} + \sigma_L \rho_{L,y}}
{\sqrt{\sigma_x^2 + \sigma_L^2 + 2 \rho_{x,L} \sigma_x \sigma_L}}\]

This is the correlation one may try to infer from the market, through historical estimation or implying it from liquid market quotes. In general the only parameters that have not been calibrated previously are $\rho_{x,y}$ and $\rho_{L,y}$. If we make for example the assumption that the two are the same,
\[ \rho_{x,y}= \rho_{L,y} =: \rho_1 , \] then we get the model correlation parameters as a function of the already calibrated parameters and of the market correlation as
\[ \rho_1 = \mbox{corr}(d \lambda_t , \ d S_t) \frac{\sqrt{\sigma_x^2 + \sigma_L^2 + 2 \rho_{x,L} \sigma_x \sigma_L}}{\sigma_x + \sigma_L } \]

\section{Swaps and counterparty risk}\label{sec:swap}

Consider now a swap contract. The propotypical swap contract is actually a portfolio of forward contracts with different maturities, and agrees on the following.

Let $t$ be the valuation time. At the future times $T_i$ in $T_{a+1}, T_{a+2}, \ldots, T_b$, a party agrees to buy from a second party a commodity at the price $K$ fixed today, on a notional $\alpha_i$. This is expressed by saying that the first party has entered a payer swap agreement. The second party has agreed to enter a receiver swap. The value of the payer commodity swap (CS) contract to the first party, at time $t$, will be
\[ \PCS_{a,b}(t,K) = \Ex_{t} \sum_{i=a+1}^b D(t,T_i) \alpha_i (S_{T_i} - K) = \sum_{i=a+1}^b \alpha_i D(t,T_i) (F(t,T_i) - K) = \sum_{i=a+1}^b \alpha_i \mbox{Fwdp}(t,T_i;K) . \]
Since the last formula is known in our oil model, in terms of the processes $x(t)$  and $L(t)$, we easily obtain a formula for the commodity swap by summation.

If we look for the value of $K$ that sets the contract price to zero, i.e. the so called forward swap commodity price $S_{a,b}(t)$, we have
\[ S_{a,b}(t) =  \frac{\sum_{i=a}^b \alpha_i D(t,T_i) F(t,T_i)}{\sum_{i=a}^b \alpha_i D(t,T_i)}    . \]

Using this rate we can also express the payer commodity swap price at a general strike $K$ as
\[ \PCS_{a,b}(t,K) = (S_{a,b}(t) - K) \sum_{i=a+1}^b \alpha_i D(t,T_i)     \]
whereas the receiver commodity swap would be
\[ \RCS_{a,b}(t,K) = (K - S_{a,b}(t)) \sum_{i=a+1}^b \alpha_i D(t,T_i)   .  \]
These formulas provide the value of these contracts when a clearing house or margining agreements are in place. However, swaps are often traded outside such contexts and as such they embed counterparty risk.

Our general formula~(\ref{generalprice}), for a payer CS, when including counterparty risk, would read in the swap case:

\begin{eqnarray}\label{generalprice:swapcom}
\mbox{PCS}^D_{a,b}(t;K) = \mbox{PCS}_{a,b}(t,K)
-&\underbrace{\lgd\,\,\mathbb{E}_t\{\mathbf{1}_{\{t
< \tau\leq T_b\}}D(t,\tau)\left(\mbox{PCS}_{a,b}(\tau,K) \right)^+}\}&\\ \nonumber
&\mbox{Positive counterparty-risk adjustment}&\\ \nonumber
= \mbox{PCS}_{a,b}(t,K)
-&\lgd\,\,\mathbb{E}_t\left\{\mathbf{1}_{\{t
< \tau\leq T_b\}}D(t,\tau)\left(\sum_{i=a+1}^b \alpha_i \mbox{Fwdp}(\tau,T_i;K) \right)^+\right\}&
\end{eqnarray}
Since the forward formula is known in our model, we can proceed similarly to the forward case to value the counterparty risk adjustment for the swap case through simulation. The receiver case is completely analogous.

%The procedure is as follows:

\section{A case study and conclusions}\label{sec:empirical}

\includegraphics{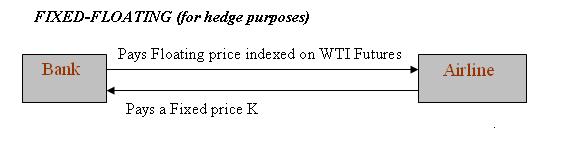}%[width=0.7\textwidth]

As a case study we consider an oil swap. An airline needs to buy oil in the future and is concerned about possible changes in the oil price. To hedge this price movement the airline asks a bank to enter a swap where the bank pays periodically to the airline a (floating) amount indexed at a relevant oil futures price at the coupon date. In exchange for this, the airline pays periodically an amount $K$ that is fixed in the beginning. \\
In the following, we are taking an example of a bank with currently high credit spreads as receiver, and one international airline as the payer of the swap.
We will look at the counterparty risk adjustment from the point of view of each of the two parties separately, by calibrating the credit model adequately in each case.

\begin{figure}[!ht]
\begin{center}
\includegraphics[width=0.685\textwidth]{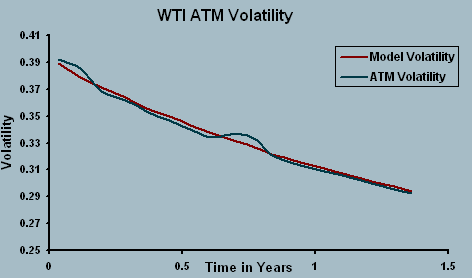}
\caption{Calibration: ATM Volatility Curve.}\label{fig:oilatmvols}
\end{center}
\end{figure}

The Oil model has been calibrated to the At The Money Futures options implied volatility shown in Figure~\ref{fig:oilatmvols}

\begin{figure}[!ht]
\begin{center}
\includegraphics[width=0.685\textwidth]{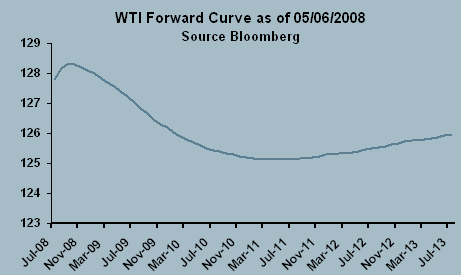}
\caption{Calibration: Forward Curve.}\label{fig:forwards}
\end{center}
\end{figure}

The shift $\varphi$ has been calibrated in order to fit the forward curve extracted from West Texas Intermediate Futures in Figure~\ref{fig:forwards}

We reformulated the commodity model by setting the parameter $\mu_L$ in the long term equilibrium price to zero, as it can be included implicitly in the deterministic shift $\varphi$.
The resulting Oil model parameters are shown in Table~\ref{tab:oilmodpar}.
\begin{table}
\begin{center}
\begin{tabular}{|cccc|} \hline
\rW $k_x$ & $\sigma_x$ & $\sigma_L$ & $\rho_{x,L}$ \\ \hline
\rW 0.7170  & 0.3522     & 0.19      & -0.0392 \\ \hline
\end{tabular}
\caption{Calibration Parameters}\label{tab:oilmodpar}
\end{center}
\end{table}

The oil swap we consider has a final maturity of 5 years, monthly payments and strike $K$ given by $K=126$ USD, that is the strike setting to zero the value of the 5 years default free oil swap.  $\alpha_i$ is equal to one (barrel).

\subsection{Counterparty Risk from the Payer Perspective (the Airline computes counterparty risk)}
First, we use the CDS spreads for the bank, which are given in Table~\ref{tab:spreadsLehman}
\begin{table}[h!]
\begin{center}
\begin{tabular}{|c|cccccc|}\hline
maturity (years) & 0.5 & 1   & 2   & 3   & 4   & 5    \\ \hline
spread (bps)     & 345 & 332 & 287 & 256 & 232 & 2.17 \\ \hline
\end{tabular}
\caption{CDS spreads term structure for the bank}\label{tab:spreadsLehman}
\end{center}
\end{table}

%The discount curve is given in Table~\ref{tab:zccurve} as a vector of continuously compound spot interest rates:\\
The yield curve is given in Table~\ref{tab:zccurve}
\begin{table}[h!]
\begin{center}
\begin{tabular}{|c|cccccc|} \hline
maturity (years) & 3/12& 6/12& 2   & 5   & 10  & 30   \\ \hline
yield (percent)  & 2.68& 2.92& 3.40& 4.27& 4.87& 5.376 \\ \hline
\end{tabular}
\caption{Zero coupon continuously compound spot interest rates}\label{tab:zccurve}
\end{center}
\end{table}

%
%\begin{itemize}
%\item \textbf{Forward}: OTC contract to buy a commodity to be delivered at a maturity date T at a price specified today. The cash/commodity exchange happens at time T.
%\item \textbf{Future}: Listed Contract to buy a commodity to be delivered at a maturity date T. Each day between today and T margins are called and there are payments to adjust the position.
%\item \textbf{Commodity Swap: Oil Example}:
%\end{itemize}

In the following we assume the bank credit quality to be characterized by a CIR++ stochastic intensity model that, as spread levels, is consistent with Table~\ref{tab:spreadsLehman} through the shift $\psi$, while allowing for credit spread volatility through the CIR dynamics.
We use the base CIR parameter set given in Table~\ref{tab:cirBase}. Later, we change the spread volatility parameter $\nu$ by reducing it through multiplicative factors smaller than one, and recalibrate the model shift to maintain consistency with Table~\ref{tab:spreadsLehman}. This way we investigate the impact of the spread volatility on the counterparty adjustment.

\begin{table}[h!]
\begin{center}
\begin{tabular}{|cccc|} \hline
$y_0$ & $\kappa$ & $\mu$ & $\nu$ \\ \hline
0.0560 & 0.6331 & 0.0293 & 0.5945 \\ \hline
\end{tabular}
\caption{CIR parameters for the base case for the bank credit spread volatility}\label{tab:cirBase}
\end{center}
\end{table}

\begin{figure}
\begin{center}
\scalebox{0.90}{
\includegraphics[width=0.8\textwidth]{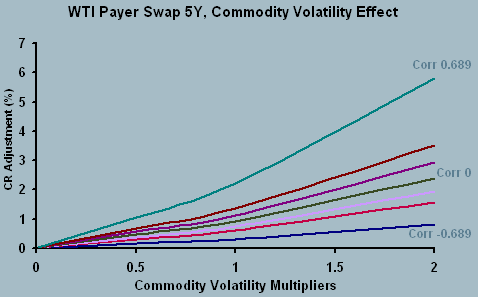}
}
\newline \bigskip\noindent {\scriptsize Fixed Leg Price maturity 5Y: 6852.35 USD for a notional of 1 Barrel per Month, CR-CVA as a (\%) of the Fixed leg price}
\caption{Commodity Swap CR-CVA Results Overview : Commodity Volatility Effect}\label{fig:cmdvolcva}
\end{center}
\end{figure}

\begin{figure}
\begin{center}
\scalebox{0.90}{
\includegraphics[width=0.8\textwidth]{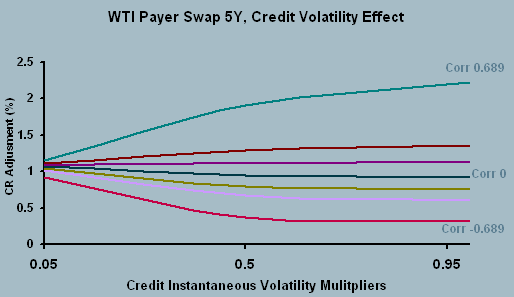}
}
\newline \bigskip\noindent {\scriptsize Fixed Leg Price maturity 5Y: 6852.35 USD for a notional of 1 Barrel per Month, CR-CVA as a (\%) of the Fixed leg price}
\caption{Commodity Swap CR-CVA Results Overview : Credit volatility effect}\label{fig:crdvolcva}
\end{center}
\end{figure}

The graphs in Fig.~\ref{fig:cmdvolcva}  and Fig.~\ref{fig:crdvolcva} illustrate some of our results for the CR-CVA. The counterparty risk is expressed as a percentage of a 5Y maturing swap fixed leg value, which is 6852.35 USD. \\
First we observe the effect of varying the commodity volatility while keeping the credit intensity volatility fixed at $\nu_{Bank} = 59\%$\footnote{The CDS implied volatility associated to these parameters is 26\%. Brigo (2005, 2006), under the CDS market model, shows that implied volatilities for CDS options can easily exceed $50\%$}
The commodity volatility was varied by applying multiplicative factors to the two factors instantaneous volatilities $\sigma_x$ and $\sigma_l$.

\begin{figure}
\begin{center}
\scalebox{0.90}{
\includegraphics[width=0.8\textwidth]{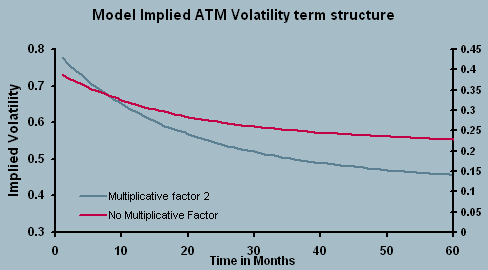}
}

\caption{Model Implied Volatility Without (right scale) and with (left scale) multiplicative factors }\label{fig:voltermstruct}
\end{center}
\end{figure}

As an indication of implied volatility levels, the term structure of the commodity implied volatility when we apply the multiplicative factor 2 is given in Fig.~\ref{fig:voltermstruct}

Secondly, we observe the effect of varying the intensity volatility while keeping the commodity spot volatility fixed at $\sigma_S = 32.82\%$ as implied by Table~\ref{tab:oilmodpar}.

The same results are presented in a different way in Tables~\ref{tab:crdvolcva} and \ref{tab:cmdvolcva}.
In these tables, we give the absolute value of the adjustment in USD. We also express it as an adjusted Strike price that the payer might choose to pay to its counterparty by taking into account the estimated adjustment:

\begin{table}
\begin{center}
\scalebox{0.90}{
\begin{tabular}{|clccc|}\hline
\rH $\bar{\rho}$ &intensity volatility $\nu_R$   & 0.0295 & 0.295  & 0.59  \\ \hline
\rW -68.9   & Adjustement in USD & 63.49  & 25.17 & 21.58  \\
\rW     & Adjusted Strike & 124.84 & 125.54 & 125.60 \\\hline
\rG -27.6   & CR-CVA (USD) & 69.99  &  45.89 & 41.5 \\
\rG         & Adjusted Strike & 124.71 & 125.16& 125.24 \\\hline
\rW -13.8   & CR-CVA (USD) & 71.83 &  55.02 & 51.48 \\
\rW         & Adjusted Strike & 124.68 & 124.99 & 125.05\\\hline
\rG 0       & CR-CVA (USD) & 73.3 &  65.23 & 63.42 \\
\rG         & Adjusted Strike & 124.66 & 124.80 & 124.84 \\\hline
\rG +13.8  & CR-CVA (USD) & 74.62  &  76.63 & 77.36 \\
\rG         & Adjusted Strike & 124.63& 124.59 &124.58 \\\hline
\rW +27.6   & CR-CVA (USD) & 75.88  &  88.93 &93.08 \\
\rW         & Adjusted Strike & 124.61 & 124.37&	124.29 \\\hline
\rG +68.9  & CR-CVA (USD) & 79.32  &  130.39 & 152.05\\
\rG         & Adjusted Strike & 124.54&	123.61&123.21 \\\hline
\end{tabular}
}

\bigskip\noindent Fixed Leg Price maturity 5Y: 6852.35 USD for a notional of 1 Barrel per Month, Fair Strike without Counterparty Risk 126 USD.
\caption{Effect of credit spread volatility on the CR-CVA}\label{tab:crdvolcva}
\end{center}
\end{table}

\bigskip

\begin{table}
\begin{center}
\scalebox{0.90}{
\begin{tabular}{|clcccc|}\hline
\rH $\bar{\rho}$ &Comdty spot vol $\sigma_S$ & 0.033 & 0.1642 & 0.3285 & 0.657 \\ \hline
\rW -68.9   & CR-CVA (USD) & 1.17   & 11.05 & 21.58 & 57.11 \\
\rW     & Adjusted Strike & 125.98&	125.79 & 125.60 &124.95 \\\hline
\rG -27.6   & CR-CVA (USD) & 1.63  &  21.75 & 41.5 & 107.48  \\
\rG         & Adjusted Strike & 125.97 & 125.60 & 125.24 & 124.03 \\\hline
\rW -13.8   & CR-CVA (USD) & 1.8  &  26.71 &51.48 & 133.49 \\
\rW         & Adjusted Strike & 125.96&	125.51&	125.05&	123.55\\\hline
\rG 0       & CR-CVA (USD) & 1.98  &  32.4  & 63.42 & 164.27 \\
\rG         & Adjusted Strike & 125.96 & 125.41 &124.84&122.98 \\\hline
\rG +13.8  & CR-CVA (USD) &  2.15 & 38.85 & 77.36 &200.08 \\
\rG         & Adjusted Strike & 125.96 &125.28 &	124.58	& 122.33 \\\hline
\rW +27.6   & CR-CVA (USD) & 2.34  &  46.05 &93.08 & 240.41 \\
\rW         & Adjusted Strike & 125.96&	125.15 &124.29&121.59 \\\hline
\rG +68.9  & CR-CVA (USD) & 2.92  &  72.47 & 152.05 & 397.87\\
\rG         & Adjusted Strike & 125.95&124.67&123.21&118.70\\\hline
\end{tabular}
}

\bigskip\noindent  Fixed Leg Price maturity  5Y: 6852.35 USD for a notional of 1 Barrel per Month, Fair Strike without Counterparty Risk 126 USD.
\caption{Effect of oil volatility on the CR-CVA}\label{tab:cmdvolcva}
\end{center}
\end{table}

\newpage

\subsection{Counterparty Risk from the Receiver Perspective (the Bank computes counterparty risk) }

Now we place ourselves from the point of view of the bank, and we use the CDS spreads for the airline, which are given in Table~\ref{tab:spreadsQuantas}
\begin{table}[h!]
\begin{center}
\begin{tabular}{|c|cccccc|}\hline
maturity (years) & 0.5 & 1  & 2   & 3   & 4   & 5   \\ \hline
spread (bps)     & 76  & 82 & 104 & 122 & 139 & 154 \\ \hline
\end{tabular}
\caption{CDS spreads term structure for the airline}\label{tab:spreadsQuantas}
\end{center}
\end{table}

We use the same discount curve as in Table~\ref{tab:zccurve}.

Here, the airline credit quality is represented by a CIR++ stochastic intensity model that, as spreads levels, is consistent with Table~\ref{tab:spreadsQuantas} through the shift $\psi$, while allowing for credit spread volatility through the CIR dynamics.
We use the base CIR parameter set given in Table~\ref{tab:cirAirline}. Later, we reduce the spread volatility parameter $\nu$ via multiplicative factors smaller than one, and recalibrate the shift to maintain each time the model consistent with Table~\ref{tab:spreadsQuantas}. This way we investigate again the impact of the spread volatility on the counterparty adjustment.

\begin{table}[h!]
\begin{center}
\begin{tabular}{|cccc|} \hline
$y_0$ & $\kappa$ & $\mu$ & $\nu$ \\ \hline
0.0000 & 0.5341 & 0.0328 & 0.2105 \\ \hline
\end{tabular}
\caption{CIR parameters for the base case for the airline credit spread volatility}\label{tab:cirAirline}
\end{center}
\end{table}

\begin{figure}
\begin{center}
\scalebox{0.90}{
\includegraphics[width=0.8\textwidth]{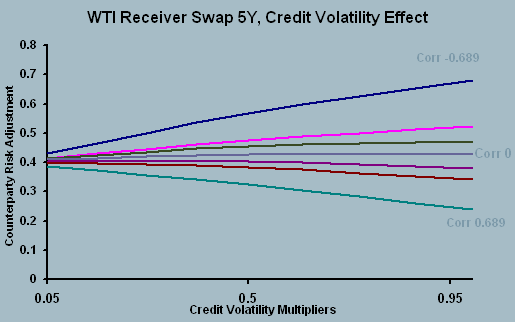}
}
\newline \bigskip\noindent {\scriptsize Fixed Leg Price maturity 5Y: 6852.35 USD for a notional of 1 Barrel per Month, CR-CVA as a (\%) of the Fixed leg price}
\caption{Commodity Swap CR-CVA Results Overview : Credit volatility effect}\label{fig:crdvolcvarec}
\end{center}
\end{figure}

\begin{figure}
\begin{center}
\scalebox{0.90}{
\includegraphics[width=0.8\textwidth]{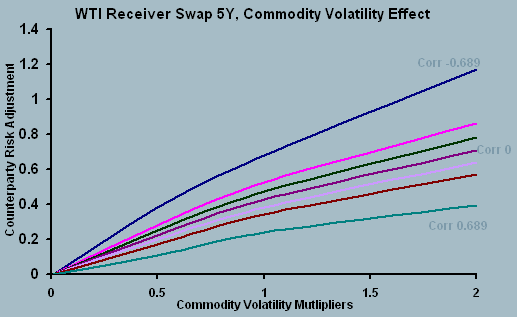}
}
\newline \bigskip\noindent {\scriptsize Fixed Leg Price maturity 5Y: 6852.35 USD for a notional of 1 Barrel per Month, CR-CVA as a (\%) of the Fixed leg price}
\caption{Commodity Swap CR-CVA Results Overview : Commodity volatility effect}\label{fig:cmdvolcvarec}
\end{center}
\end{figure}

As before, we observe the effect of varying the commodity volatility and of the airline credit intensity volatility, starting from $\sigma_S = 32.82\%$ as from Table~\ref{tab:oilmodpar} and $\nu_{Airline} = 21\%$. We apply the same multiplicative factors as before and the results are summarized in the graphs in Fig.~\ref{fig:cmdvolcvarec}  and Fig.~\ref{fig:crdvolcvarec}.

The same results are presented more in detail in Tables~\ref{tab:crdvolcvarec} and \ref{tab:cmdvolcvarec}.

\begin{table}
\begin{center}
\scalebox{0.90}{
\begin{tabular}{|clccc|}\hline
\rH $\bar{\rho}$ &intensity volatility $\nu_R$   & 0.0295 & 0.295  & 0.59  \\ \hline
\rW -68.9   & CR-CVA (USD) & 29.62  & 38.95 & 46.62  \\
\rW     & Adjusted Strike &126.54&126.71&126.85\\\hline
\rG -27.6   & CR-CVA (USD) & 28.41  &  32.58  & 35.82 \\
\rG         & Adjusted Strike & 126.52&	126.59&126.66\\\hline
\rW -13.8   & CR-CVA (USD) & 28.21 & 31.02  & 32.4   \\
\rW         & Adjusted Strike &126.52&	126.57&	126.59\\\hline
\rG 0       & CR-CVA (USD) & 27.99 &  29.37 & 29.16 \\
\rG         & Adjusted Strike &  126.51&126.54&	126.53 \\\hline
\rG +13.8  & CR-CVA (USD) & 27.78  & 27.72  & 26.09 \\
\rG         & Adjusted Strike & 126.51&	126.51&126.48 \\hline
\rW +27.6   & CR-CVA (USD) & 27.49& 26.15  & 23.42 \\
\rW         & Adjusted Strike & 126.50&	126.48&	126.43 \\\hline
\rG +68.9  & CR-CVA (USD) & 26.48  &  22.23 & 16.31\\
\rG         & Adjusted Strike & 126.48&	126.41&	126.30 \\\hline
\end{tabular}
}

\bigskip\noindent Fixed Leg Price maturity  5Y: 6852.35 USD for a notional of 1 Barrel per Month, Fair Strike without Counterparty Risk 126 USD.
\caption{Effect of credit spread volatility on the CR-CVA}\label{tab:crdvolcvarec}
\end{center}
\end{table}

\bigskip

\begin{table}
\begin{center}
\scalebox{0.90}{
\begin{tabular}{|clcccc|}\hline
\rH $\bar{\rho}$ &Comdty spot vol $\sigma_S$ & 0.033 & 0.1642 & 0.3285 & 0.657 \\ \hline
\rW -68.9   & CR-CVA (USD) & 0.12   &26.33 & 46.62 & 80.26 \\
\rW     & Adjusted Strike & 126.00&	126.48&	126.85&	127.47\\\hline
\rG -27.6   & CR-CVA (USD) & 0.09  & 19.33& 35.82&59.23  \\
\rG         & Adjusted Strike & 126.00&126.35&	126.65&	127.08 \\\hline
\rW -13.8   & CR-CVA (USD) & 0.08  & 17.35& 32.4& 53.64 \\
\rW         & Adjusted Strike & 126.00&126.32&	126.59&	126.98\\\hline
\rG 0       & CR-CVA (USD) & 0.07  & 15.42   & 29.16 & 48.59 \\
\rG         & Adjusted Strike &  126.00&126.28&	126.53&	126.89 \\\hline
\rG +13.8  & CR-CVA (USD) & 0.06  & 13.58 & 26.09 & 43.88\\
\rG         & Adjusted Strike & 126.00&	126.25&	126.48&	126.80 \\\hline
\rW +27.6   & CR-CVA (USD) & 0.05  &  11.86 &23.42 & 39.09 \\
\rW         & Adjusted Strike & 126.00&	126.22&126.43&126.72 \\\hline
\rG +68.9   & CR-CVA (USD) & 0.03   & 7.4 & 16.31 & 27.16 \\
\rG         & Adjusted Strike & 126.00&	126.13&	126.30&126.50 \\\hline
\end{tabular}
}

\bigskip\noindent  Fixed Leg Price maturity  5Y: 6852.35 USD for a notional of 1 Barrel per Month, Fair Strike without Counterparty Risk 126 USD.
\caption{Effect of oil volatility on the CR-CVA}\label{tab:cmdvolcvarec}
\end{center}
\end{table}

\newpage

\subsection{Conclusions}
The patterns we observe in the counterparty-risk credit valuation adjustment (CR-CVA) are natural.
Starting with the receiver case, for a fixed credit spread volatility, the receiver CR-CVA increases in oil volatility and decreases in correlation. Given the embedded oil option, the increase with respect to oil volatility is natural (as is in the payer case). As concerns correlation, as this increases, the oil tends to move in line with credit spreads. This means that higher credit spreads will lead to higher oil values, and the option will end up less in the money as the oil spot goes up. The opposite appears in the payer case. Patterns in credit spread volatility are similarly explained.

The size of the CVA hence depends on the precise value of the volatility and correlation dynamic parameters that cannot be explained via rough multipliers.

\end{document}